# Phase Transition and Magnetic Suppression in Heusler Alloy IrMnAl

*Himanshu Joshi[1*], Amel Laref[2], A. Yvaz[3] and Dibya. P. Rai[4*]*

[1] Department of Physics, SRM University Sikkim, Gangtok-737102, India

[2] Department of Physics and Astronomy, College of Science, King Saud University, Riyadh-11451, Saudi Arabia

[3] World-class Research centre "Advanced Digital Technologies", State Marine Technical University, Saint Petersburg – 190121, Russia

[4] Department of Physics, Mizoram University, Aizawl-796004, India

**E-mail:** himanshuabijoshi09@gmail.com (H. Joshi); dibyaprakashrai@gmail.com (D.P. Rai)

**Abstract:** This study conducts a comprehensive first-principles investigation of the IrMnAl Heusler alloy, highlighting its magnetic properties and assessing the effects of pressure-induced phase shift. The limitations of Generalized Gradient Approximation in accurately representing the compound's magnetic behavior is addressed by employing the GGA+U approach, which more effectively captures the electronic features. At 5.6 GPa pressure, we observe a novel structural phase transition from the cubic $F\bar{4}3m$, to the $P\bar{4}3m$ space group. The calculated Curie temperature, combined with the analysis of Mn-Mn exchange interaction, sheds light on the weak ferromagnetism observed in the new phase of the compound.

## 1. Introduction

Mn-based half-Heusler compounds have recently gained significant interest primarily due to their remarkable magnetic properties, particularly their ferromagnetic characteristics [1]. These compounds demonstrate half-metallicity, with the majority spin characterised as a conductor, while the minority spin behaves as a semiconductor, highlighting their unique electronic structure [2]. Such half-metallic ferromagnets (HMF) play a vital role in generating a highly spin-polarized current, which is particularly important for advanced applications in ultrahigh-density memory technology [3]. Half-metallicity in Heusler alloys depends upon the cumulative magnetic moments that follows from the Slater-Pauling rule [4] and usually Mn- based Heusler are also expected to show HMF properties. However, studies on the electronic properties of IrMnAl points to the missing HMF behaviour [1], characterized by metallic nature of both the majority and minority spin, observed from their band structure.

IrMnAl is identified as weak ferromagnet, with Curie temperature in the range of 400 K temperature [5], a value surprisingly close to that of other Mn-based HMFs. This similarity in Curie temperature makes IrMnAl an intriguing subject for further study, particularly in understanding why it does not exhibit the expected HMF behavior. Analyzing the electronic and magnetic properties of IrMnAl could provide valuable insights into its unique characteristics. However, the literature on the detailed electronic structure of this compound is limited, likely due to its non-HMF nature. A similar suppression of HMF has been observed in Mn-based NiMnSb [6], a phenomenon often attributed to atomic disorder resulting from the interchange of Ni and Mn vacancies [7]. Such atomic disorder is not uncommon in Heusler alloys [8-10] and can significantly influence the material's electronic properties. In IrMnAl, this characteristic disorder arises from Ir and Al atoms arbitrarily occupying the 8c orientation within the C1 structure in which the compound crystallizes. Other Mn-based Heusler alloys are reported to lose HMF property due to anti-site defects which induce minority gap states and further complicate the materials electronic behaviour [11, 12].

Recent studies suggest that GGA formalism may not be sufficient for accurately capturing the electronic configuration of HMFs [13]. The true half-metallic nature of certain Heusler compounds, such as $Co_2FeGe$, $Co_2MnSi$, and $Co_2FeSi$, has only been correctly identified with the introduction of Hubbard-Coulomb potential 'U' [14, 15]. Therefore, to address site-specific correlations at the transition metal positions, we also employed the GGA + U approach [16] in this study. Unlike pure LDA or GGA, the + U approach incorporates an orbital-specific treatment of Coulomb and exchange interactions, providing a more accurate representation of the electronic structure. Specifically, the effective interaction between Coulomb and exchange ($U_{eff} = U−J$) has been applied in our calculation to address the limitations of the LSDA or GGA methods. Generally, the addition of U to the LDA or GGA framework results in increased band gap width. However, in the case of IrMnAl, the metallic nature in both the spin channels sustained, even after the inclusion of U. It is crucial to recognize that incorporating U does not induce half-metallic ferromagnetism but rather affirms the half-metallic nature of these compounds and aids in accurately explaining magnetic properties. Additionally, the local correlation affects the electronic states by increasing the separation between bands of varying symmetries, which in turn leads to a shift in the Fermi energy relative to the energy difference in the minority states, or vice versa.

Another subsequent approach involves investigating whether the material's characteristics change under pressure. Applying pressure can potentially alter the atomic arrangement and electronic structure, which may reduce or eliminate the aforementioned anti-site defects or atomic disorder that cause the loss of half-metallic properties. Various pressure dependent studies on half-metallicity of Heusler alloys already exists [17-20]. Nonetheless, the metallic property of IrMnAl remained dominant under pressures reaching up to 6 GPa, past which the compound is forecasted to lose stability. Interestingly, at 5.6 GPa pressure, a transistion in phase was observed in IrMnAl, a phenomenon not previously reported for this compound. This phase transition occurs within the cubic crystal symmetry but involves a change in space group from $F\bar{4}3m$ to $P\bar{4}3m$. We refer to the IrMnAl structure in $F\bar{4}3m$ as the α-phase and in $P\bar{4}3m$ as the β phase. The transition results in reduction of translational symmetry due to the loss of face-centering, as well as a decrease in rotational symmetry with the disappearance of fourfold axes [21]. Additionally, there are fewer mirror planes and a loss of inversion symmetry. These changes lead to a simpler, lower-symmetry structure in the $P\bar{4}3m$ phase, which impact the material's physical properties, including its electronic structure.

Motivated by these findings, we conduct extensive analysis on the electronic behaviour of the β-phase, along with X-ray Magnetic Circular Dichroism (XMCD) to gain a deeper understanding on the magnetic properties. Furthermore, we explore the temperature dependent magnetization and the Mn-Mn exchange interaction, which helped to gain further insight into the material's weak ferromagnetism. This study presents the primary evidence of phase transition in Mn-based Heusler alloys and provides novel understanding of the P$\bar{4}$3m phase in IrMnAl.

## 2. Computational Details

Creating and refining new, advanced materials in a laboratory setting is a complex, time-consuming, and expensive endeavour that requires extensive testing and costly prototypes. To streamline this process, ab initio calculation methods have become indispensable. Given the increasing importance of computational approaches, density functional theory (DFT) [22, 23] based calculations form the foundation of this work. For this study, we utilized the Wien2k software [24], which employs the full-potential linearized augmented plane wave (FP-LAPW) [25] method. Ground state features were computed using the Perdew-Burke-Ernzerhof (PBE) generalized gradient approximation (GGA) [26]. It's well-documented that PBE tends to underestimate band gap values, a limitation that has been consistently reported in many ab initio studies [27-29]. The underestimation of the gap in GGA is primarily owing to the impractical self-Coulomb repulsion it accounts for. One approach to address this issue is by correcting the self-Coulomb interaction anomaly using precise Hartree-Fock exchange, though this method is computationally expensive. A more efficient alternative is the GGA+U method [30], which is less demanding computationally and involves the assessment of the onsite Coulomb self-interaction parameter (U) to mitigate the error in self-Coulomb repulsion, serving as a correction over GGA. The effective Coulomb potential is obtained using $U_{eff} = U_{Ir/Mn} - J_{Ir/Mn}$, with $J_{Ir} = 0$ eV [31, 32]. The electronic structure showed negligible variation within $U_{Ir}$=1.0 eV and $U_{Mn}$= 3.4 – 4.1 eV, therefore the interaction parameter was taken as U$_{Ir}$ ≈ 1.0 eV and U$_{Mn}$ ≈ 3.67 eV. Within the muffin tin (MT) spheres, charge density and potential contributed from non-spherical components are considered extending to maximum angular momentum value of $l_{max}$ = 10. The threshold limit is set as $R_{MT} \times K_{max} = 7$, where $K_{max}$ represents the largest reciprocal lattice vector in the plane wave expansion and $R_{MT}$ the smallest atomic radius among all atoms. The Fourier expansion of the charge density and potential describes their extension to the interstitial region, with wave vectors not exceeding G$_{MAX}$ = 12 a.u $^{-1}$. To ensure self-consistency, the energy convergence criterion was established at $10^{-5}$ Ry, and k-mesh of 20 × 20 × 20 was used within the first Brillouin zone. Experimental analysis of IrMnAl have confirmed the presence of magnetic ordering in these materials. Consequently, spin polarised simulations were conducted using lattice constants of 5.99 Å, derived from experimental data, so as to maintain compatibility with experimental results.

## 3. Results

### 3.1. Structural properties and Phase transition of the emerging phase

The α-phase of IrMnAl exhibits a crystal structure characteristic of the AlLiSi type, identified by the space group F$\bar{4}$3m (Figure 1a). The Wyckoff positions are: Ir is located at the 4*b* site, Mn occupies the 4*a* site, and Al is positioned at the 4*d* site [5]. To explore the possible phase transition, we employ a 2×2×2 supercell of the α-phase, which at a pressure of 5.6 GPa (Figure 1d), changes into the β-phase with space group P$\bar{4}$3m (Figure 1b). The new phase remains within the cubic crystal class, with calculated lattice constant of 5.873 Å, which is approximately 2% lower than that of α-phase. Figure (1c) presents the optimization curves of α- and β-phase, both fitted using the Murnaghan's equation of state [33], revealing that the α-phase exhibits greater energetic stability and is more frequently encountered in natural settings. At the transition pressure, the volumes of the α- and β-phase are expected to be 214.92 Å$^3$ and 202.57 Å$^3$ respectively. The enthalpy plot between the two phases as a function of pressure (Figure 1d) confirms the phase transition. The crystal structure details of the two phases are tabulated in Table 1.

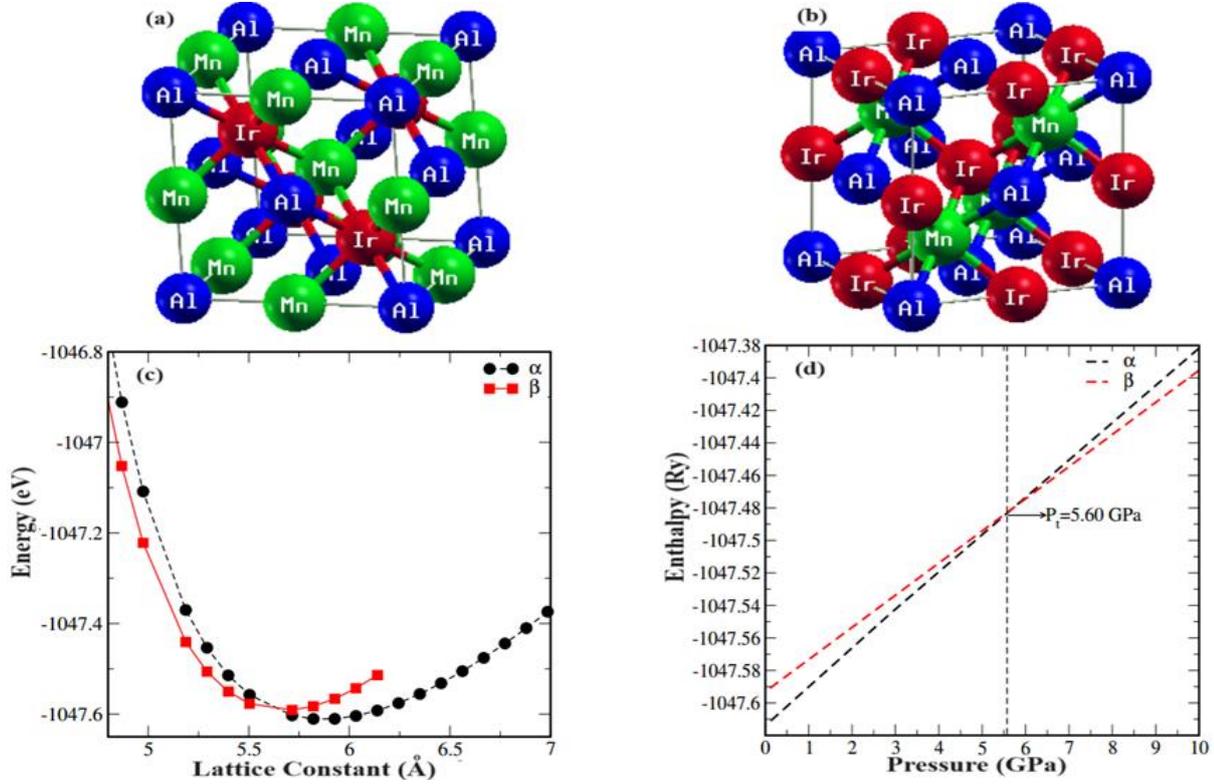

**Figure 1:** **(a)** The crystal structure of IrMnAl in $F\bar{4}3m$, **(b)** the crystal structure of IrMnAl in $P\bar{4}3m$, **(c)** variation of total energy with lattice constant, and **(d)** phase transition from α-IrMnAl to β-IrMnAl, with the vertical dashed line indicating the transition pressure.

**Table 1:** Optimized crystal structure details of the two phases

| Phase | Space group | Space Group No. | Crystal class | a (Å) (GGA) | a (Å) (GGA+U) |
|---|---|---|---|---|---|
| α-phase | F43m | 216 | f.c.c | 5.92 | 6.01 |
| β-phase | P43m | 215 | s.c | 5.87 | 5.93 |

### 3.2. Elastic Properties

The elastic properties were computed using Density Functional Theory (DFT) within the context of Lagrangian elasticity, where the material is modelled as an elastic medium characterized by anisotropy and homogeneity. Second-order elastic parameters were derived through energy-strain approach, utilizing the ElaStic code [34]. In the cubic structure studied, three distinct elastic constants were identified: $C_{11}$, $C_{12}$, and $C_{44}$, as described by the elastic tensor -

$$\begin{pmatrix} C_{11} & C_{12} & C_{12} & 0 & 0 & 0 \\ C_{12} & C_{11} & C_{12} & 0 & 0 & 0 \\ C_{12} & C_{12} & C_{11} & 0 & 0 & 0 \\ 0 & 0 & 0 & C_{44} & 0 & 0 \\ 0 & 0 & 0 & 0 & C_{44} & 0 \\ 0 & 0 & 0 & 0 & 0 & C_{44} \end{pmatrix}$$

These constants were then employed to calculate various elastic properties using the averaging methods proposed by Voigt, Reuss, and Hill [35, 36, 37]. The various modulus of elasticity were then calculated as follows-

$$\text{Bulk Modulus } (B) = \frac{1}{3}\left(C_{11} + 2C_{12}\right) \quad (1)$$

The bulk modulus for a cubic structure is the same when calculated using the Voigt, Reuss, and Hill averaging methods. The shear modulus ($G$) for these methods is specified by equation 2, 3, and 4, respectively-

$$G_V = \frac{(C_{11} - C_{12} + 3C_{44})}{5} \quad (2)$$

$$G_R = \frac{5(C_{11} - C_{12})C_{44}}{4C_{44} + 3(C_{11} - C_{12})} \quad (3)$$

The Hill shear modulus is calculated as the average of the Voigt and Reuss shear moduli.

$$G_H = \frac{G_V + G_R}{2} \quad (4)$$

The Poisson's ratio ($\eta$) and Young's modulus ($Y$) are determined using equations (5) and (6) as follows –

$$\eta = \frac{3B - 2G}{2(3B + G)} \quad (5)$$

$$Y = \frac{9BG}{3B + G} \quad (6)$$

By substituting G with $G_V$ and $G_R$ in equation (5) and (6), the Voigt and Reuss averages for Young's modulus and Poisson's ratio can be calculated.

In relation to the compressional ($v_l$) and shear ($v_s$) sound velocities, the mean sound velocity ($v_m$) is represented as

$$v_m = \frac{1}{3}\left(\frac{2}{v_s^3} + \frac{1}{v_l^3}\right)^{-\frac{1}{3}} \quad (7)$$

Equation 8 and 9 describe the longitudinal ($v_l$) and shear ($v_s$) wave velocities in relation to the material density ($\rho$).

$$v_l = \sqrt{\frac{3B + 4G}{3\rho}} \quad (8)$$

$$v_s = \sqrt{\frac{G}{\rho}} \quad (9)$$

The α-phase exhibits minimal variation in the calculated elastic moduli between the GGA and GGA+U methods, shown in Table 2, indicating that electron correlation effects are less significant in this phase. In contrast, the β-phase shows a marked deviation between the GGA and GGA+U results, likely due to the stronger electron correlation effects in this phase. These effects are more accurately captured by the GGA+U method, leading to the observed differences. This indicates that the band structures calculated using GGA and GGA+U are expected to differ notably in the β-phase, as evidenced by the band structure plots presented later in the manuscript. Also, the estimated bulk modulus for the β-phase is larger, indicating high resistance to volume deformation in comparison to the α-phase. However, due to significant variations between the GGA and GGA+U schemes, it is difficult to conclusively determine which phase exhibits a higher ability to withstand shape deformation, associated with the shear modulus (G), or the material's response to elongation and compression, related to Young's modulus (Y). The high elastic wave velocities in both phases indicate a tightly packed arrangement of atoms within their crystal structures.

**Table 2:** Calculated elastic moduli of the α- and β-phase using GGA and GGA+U methods

| Moduli of Elasticity | α-phase | | β-phase | |
|---|---|---|---|---|
| | GGA | GGA+U | GGA | GGA+U |
| $C_{11}$ (GPa) | 165.327 | 165.933 | 248.882 | 152.301 |
| $C_{12}$ (GPa) | 103.444 | 95.729 | 123.002 | 176.696 |
| $C_{44}$ (GPa) | 70.174 | 67.554 | 115.971 | 92.00 |
| B (GPa) | 124.072 | 119.130 | 164.962 | 168.565 |
| $Y_V$ (GPa) | 142.575 | 142.032 | 238.592 | 137.432 |
| $G_V$ (GPa) | 54.481 | 54.574 | 94.7759 | 50.374 |
| $Y_R$ (GPa) | 124.150 | 130.010 | 221.409 | 123.453 |
| $G_R$ (GPa) | 46.560 | 49.317 | 86.738 | 38.054 |
| $Y_H$ (GPa) | 133.363 | 136.021 | 230.000 | 130.25 |
| $G_H$ (GPa) | 50.521 | 51.945 | 90.749 | 44.2 |
| $\eta_V$ | 0.31 | 0.30 | 0.26 | 0.36 |
| $\eta_R$ | 0.33 | 0.32 | 0.28 | 0.62 |
| $\eta_H$ | 0.32 | 0.31 | 0.27 | 0.49 |
| $v_l$ (m/s) × $10^4$ | 4.675 | 4.739 | 5.302 | 4.380 |
| $v_s$ (m/s) × $10^4$ | 2.401 | 2.488 | 2.987 | 0.818 |
| $v_m$ (m/s) × $10^4$ | 3.764 | 3.768 | 4.027 | 4.227 |

### 3.3. Electronic and Magnetic Properties

The selection of GGA and GGA+U potentials significantly influence the electronic structure, especially in the β-phase, as evident from the DOS and energy band plots. The total DOS plots presented in Figure 2 reveal metallic characteristics across both spin channels for both phases. Additionally, there is notable shift in the DOS peaks when comparing the results obtained from the two computational schemes. The Density of States plot is divided into several key regions within the valence and conduction bands: Region I: Extending from 0 to -2 eV, Region II: Spanning from -2 to -6 eV, Region III: Ranging from -6 to -8 eV. The conduction region is taken as region IV (0 to 4 eV) and region V (4 eV to 8 eV). The region below -8 eV is considered an intermediate zone and does not contribute to the valence properties. For the α-phase, all dominant DOS peaks are concentrated in Region II for the spin-down channel, while in the spin-up channel, DOS peaks are distributed across all three regions.

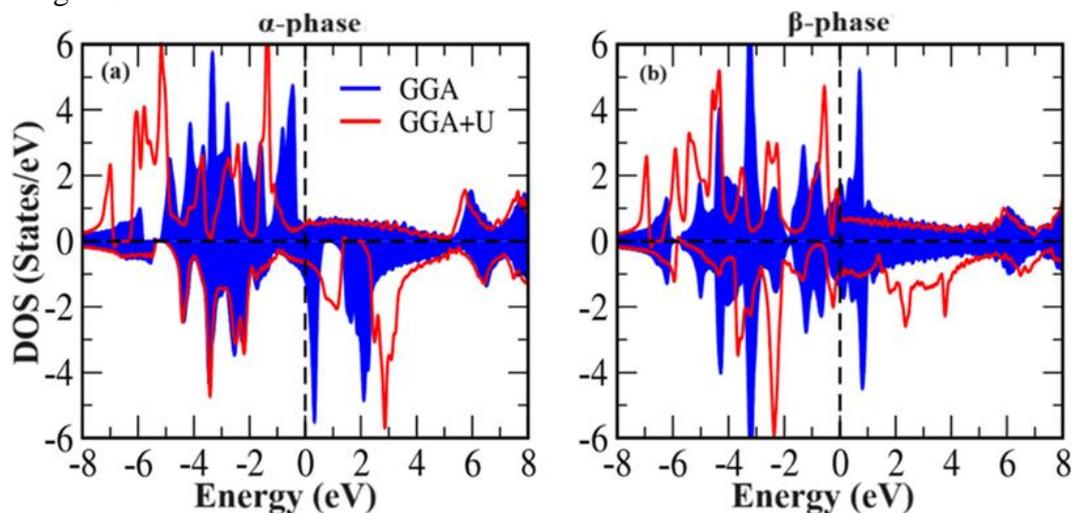

**Figure 2:** Total DOS of IrMnAl, plotted using GGA and GGA+U for the **(a)** the α-phase and **(b)** the β-phase.

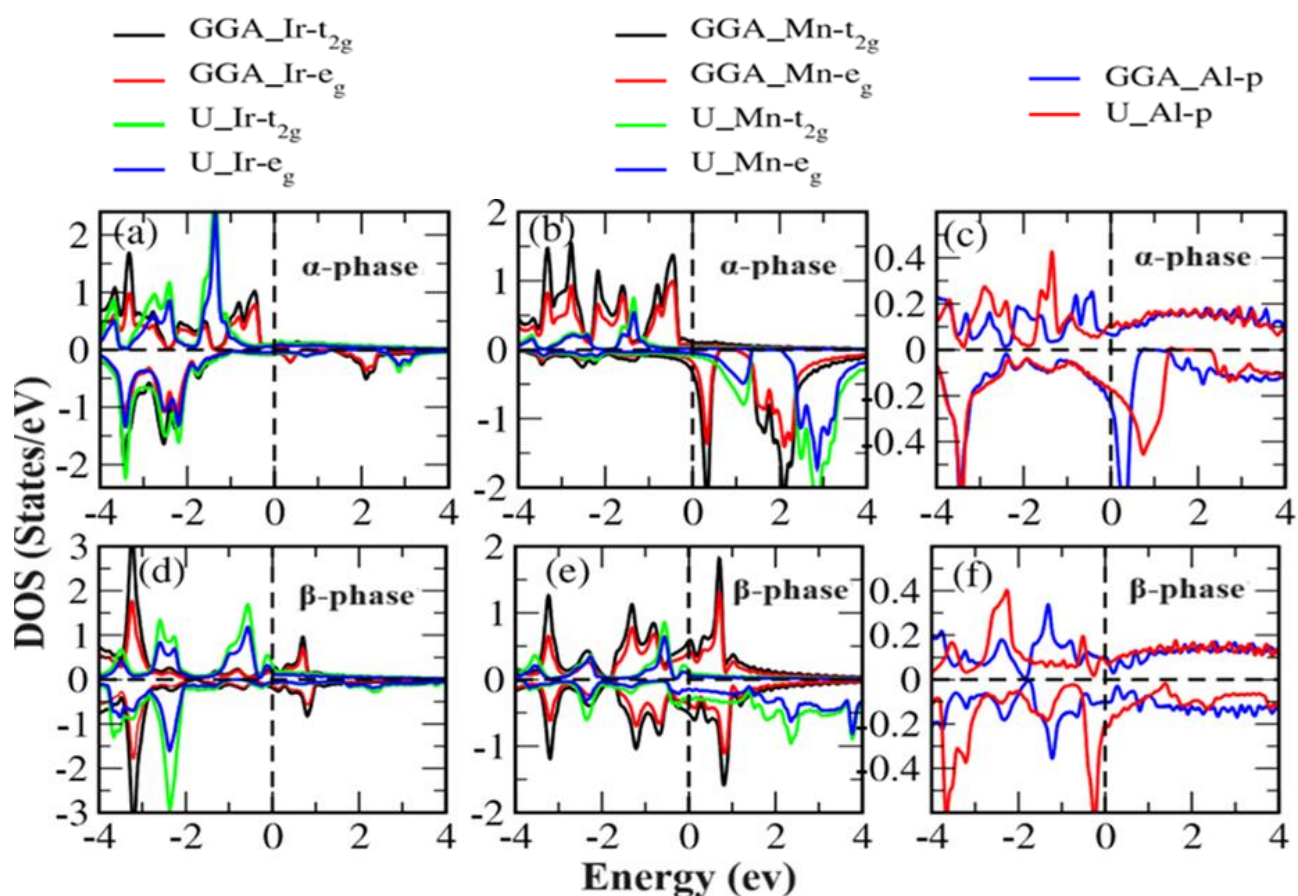

**Figure 3:** Partial Density of States for the α-phase and β-phase using GGA and GGA+U. Panels **(a)**, **(b)**, and **(c)** show the PDOS for Ir, Mn, and Al atoms in the α-phase, while panels **(d)**, **(e)**, and **(f)** display the PDOS for the same atoms in the β-phase.

The conduction band exhibits a sharp peak only in the spin-down channel. In the β-phase, similar characteristics are observed, although the conduction peaks are somewhat reduced. The inclusion of on-site Coulomb interactions, as shown in Figure 2, leads to a noticeable shift in the DOS peaks. GGA often underestimates the electron-electron repulsion in localized states, which can result in inaccurate energy levels. The U term, however, enhances the handling of correlation effects among electrons in partially filled orbitals, leading to a more precise distribution of electronic states and, consequently, a shift in the DOS peaks.

The PDOS analysis extending between the energy of -4 eV to 4 eV provides insight into the electronic structure of both the α- and β-phase. In the α-phase spin-up channel, figure 3 (a-c), the decrease in band gap is linked to the hybridization between Ir and Mn $d$-states, along with a notable contribution from Al $p$-states. This interaction causes the states to draw closer in energy, effectively destroying the gap. Conversely, the negative spin channel retains its metallic nature, as indicated by the overlap of Mn $d$-states and Al $p$-states. The observed splitting of $d$-states into $e_g$ doublets and $t2_g$ triplets in both Ir and Mn atoms is indicative of crystal field effects, as shown by the distinctive features in the PDOS. In the β-phase, as illustrated in Figure 3 (d-f), the narrowing of the band gap is similarly attributed to the strong interaction between Al p-states and Mn d-states, mirroring the behaviour seen in the α-phase. The primary impact of the GGA+U method is a shift in the DOS peaks, without major alterations to the electronic structure, which is evident from the comparison between GGA and GGA+U results in both phases. These observations are in line with previously reported studies [1], lending further credibility to the current findings.

The band structure plotted within the GGA and GGA+U confirms the metallic characteristic, for both the α-phase, as well the β-phase (Figure 4). A defining aspect of the IrMnAl electronic configuration is the significant hybridization between Mn and Ir states, with the Ir states exhibiting greater delocalization. This hybridization induces magnetism in the otherwise nonmagnetic Ir atoms within the alloy. The magnitude of the magnetic moment for Ir varies depending on the quantity of neighbouring Mn atoms and the distances between Mn and Ir atoms. Experimental findings revealed that IrMnAl exhibits a notably low net magnetic moment with the Mn site contributing only a minimal magnetic moment. Consequently, magnetization measurements have categorized IrMnAl as exhibiting weak ferromagnetism with very low overall magnetization [5]. XMCD spectroscopy, which leverages synchrotron radiation, has been established as a valuable technique for investigating magnetic moments specific to individual elements in ferromagnetic materials [38, 39]. While extensive literature exists on XMCD results for the α-phase of IrMnAl [1, 2], revealing a weak and oppositely signed magnetic moment associated with Ir due to strong Ir-Ir hybridization, this study presents findings for the β-phase.

X-ray absorption spectroscopy (XAS) offers information on how materials absorb X-rays as their energy levels changes. A peak appears in the absorption spectrum when the energy of the X-rays corresponds to the energy needed to promote an electron from a core level to a higher, unoccupied state. For Ir (Figure 5a), the XAS data at the L$_2$ and L$_3$ edges correspond to transitions from the core $2p$ states to the $5d$ conduction band. The inclusion of U results in slight shifts and broadening of these XAS peaks, suggesting modifications in the electronic structure of the $5d$ Ir-states. The XMCD plots for the same L$_3$ and L$_2$ edges provide information on the magnetic moments, with notable differences in intensity between the 'without U' (black) and 'with U' (red) cases. The inclusion of U alters the XMCD intensity, indicating changes in the magnetic properties of the Ir atoms, particularly their spin and orbital magnetic moment contributions.

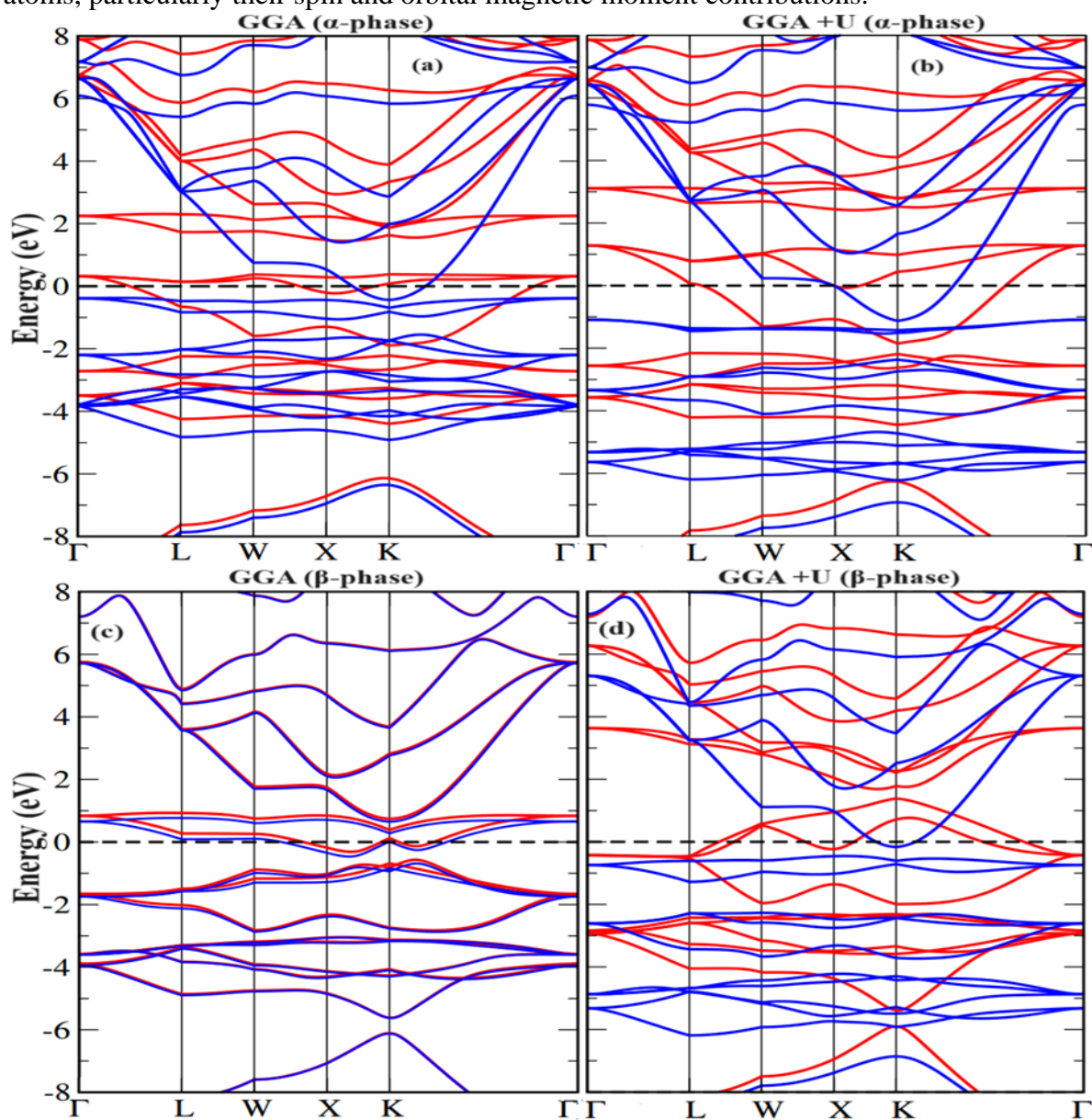

**Figure 4**: Band structure of IrMnAl plotted using GGA **(a)**, **(c)** and GGA+U **(b)**, **(d)** approach of the two phases. The red lines correspond to spin up channel and blue the spin down channel.

Similarly, the XAS peaks for Mn (Figure 5b) correspond to transitions from the core Mn *2p* state to the Mn *3d* states. The inclusion of U causes only minor shifts and slight changes in peak intensity, indicating a subtle impact on the Mn *3d* electronic structure. The effect of U on the XMCD spectra for Mn is less pronounced than for Ir, suggesting that the U correction has a limited impact on the magnetic behaviour of Mn. Overall, the shifts in the XAS peaks and changes in the XMCD spectra due to U inclusion indicate alterations in the electronic and magnetic moments, potentially enhancing the contributions from both spin and orbital components. However, as seen in the α-phase, the Ir magnetic moment remains significantly lower than that of Mn in the β-phase as well. The very small magnetic moment associated with the Ir atom results in weak magnetic moment in β-IrMnAl compound.

The decrease in magnetic moments in IrMnAl can be ascribed to the notable hybridization among Al atom *sp*-band conduction electrons, Mn *3d* electrons, and Ir *5d* electrons. The Al atoms, which likely exhibit a near 5⁻ valence state, mainly provide *sp*-character conduction electrons in the compound. These *sp*-electrons from Al interact with the *d*-electrons in an antiferromagnetic manner, leading to the destabilization of the local magnetic moment on Mn. Consequently, this antiferromagnetic coupling results in a substantial suppression of the Mn local moment [2]. Nevertheless, a small portion of Mn *3d*, which remain partially delocalized, still develop spin polarization and order ferromagnetically. The small spin moment in Ir, combined with the low overall moment in IrMnAl, is associated with a noncollinear arrangement of the Mn moments within the material [1]. In this scenario, Mn atoms may possess a significantly higher magnetic moment compared to the overall moment, due to the noncollinear magnetic arrangement in the ferromagnetic state of IrMnAl.

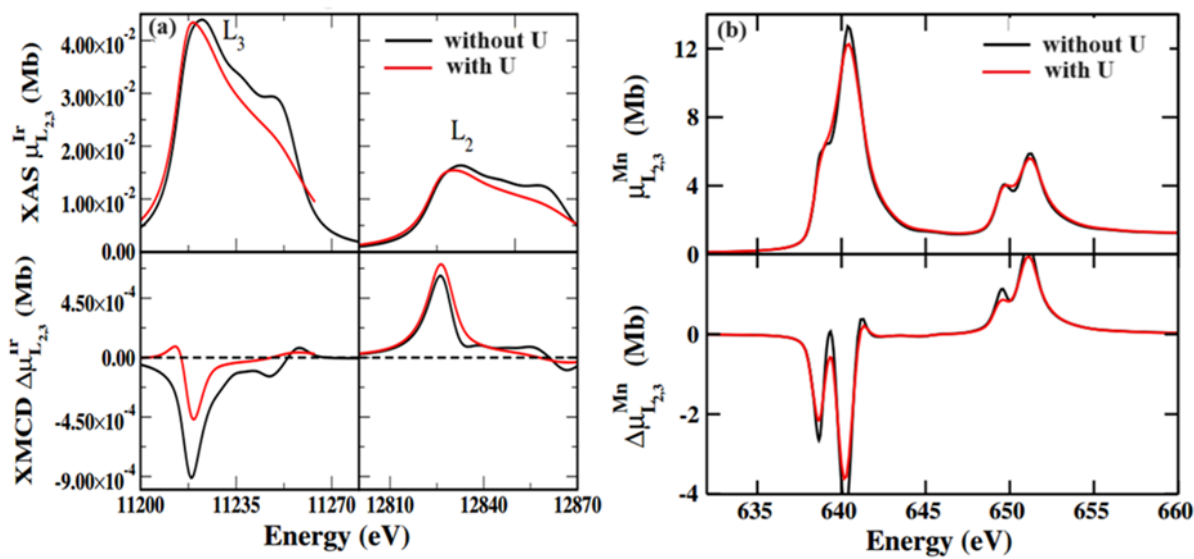

**Figure 5:** X-ray Absorption Spectra (XAS) and X-ray Magnetic Circular Dichroism (XMCD) spectra for **(a)** Ir and **(b)** Mn atoms of IrMnAl, with and without the inclusion of the Hubbard U correction, in the β-phase.

The figure below offers a closer look at the Mn-Mn exchange interactions in the IrMnAl compound, depicted relative to the separation between Mn atoms (Figure 6a). This graph showcases the interactions among various Mn pairs—Mn1-Mn2, Mn1-Mn3, Mn1-Mn4, Mn2-Mn3, Mn2-Mn4, and Mn3-Mn4, each distinguished by specific symbols and colours. At shorter distances, approximately 5 a.u., certain pairs like Mn1-Mn3, Mn1-Mn4, and Mn3-Mn4 exhibit strong antiferromagnetic interactions, as evidenced by their negative $J_{ij}$ values. Meanwhile, the Mn2-Mn3 interaction, initially weakly ferromagnetic at these shorter distances, diminishes considerably as the separation increases. Beyond 10 a.u., most interactions stabilize near zero, underscoring the short-range nature of these exchange interactions. This pattern is key to understanding the magnetic coupling in IrMnAl, which plays a significant role in defining its magnetic structure and ordering. To further explore the magnetic properties, we plotted the magnetization $\mu$, normalized to the saturation magnetization $\mu_s$, against temperature (Figure 6d). As the temperature rises, the magnetization decreases, eventually reaching zero at the Curie temperature (TC), which we calculated to be 243.4

K for the β-phase of IrMnAl. This new phase exhibits a Curie temperature approximately 39% lower than that of the lower energy phase.

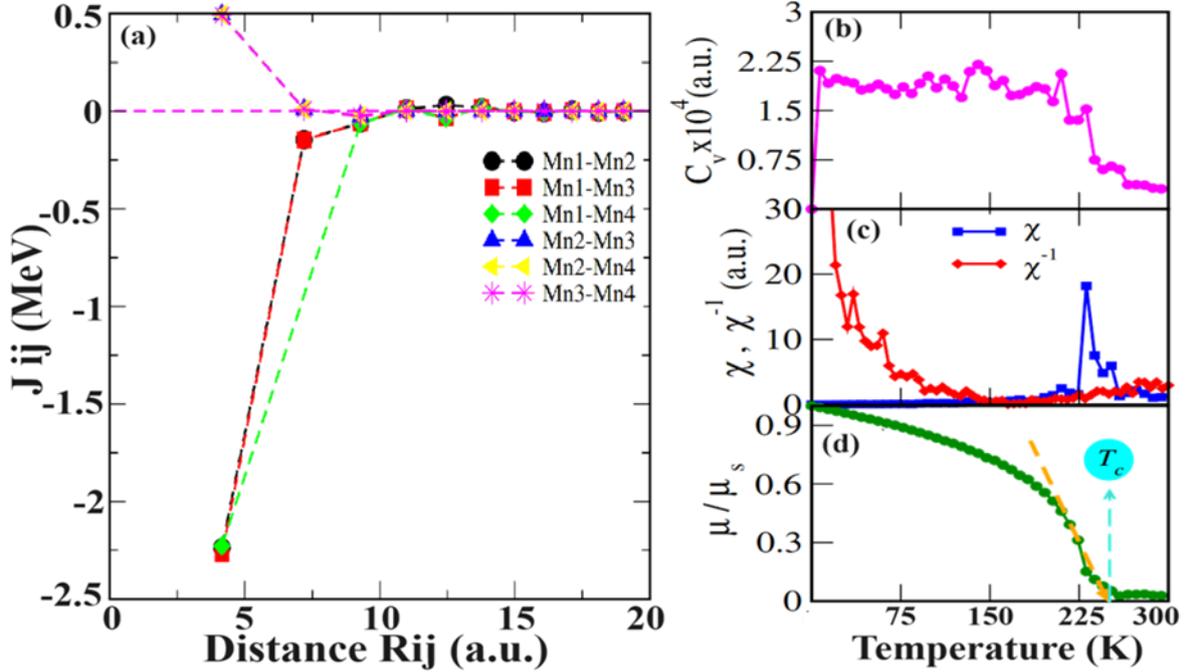

**Figure 6:** **(a)** Mn-Mn exchange interactions in IrMnAl, plotted in relation to the distance between Mn atoms, **(b)** specific heat ($C_v$), **(c)** magnetic susceptibility ($\chi$) and inverse susceptibility ($\chi^{-1}$), **(d)** magnetization plotted as a function of temperature.

Beyond this threshold, the β-phase loses its magnetic configuration, predominantly maintained by the Mn atoms, and shifts into a random paramagnetic state. Figure 6(a) illustrates the specific heat ($C_v$) variation with temperature, remaining relatively stable until it nears the Curie temperature, where a significant drop occurs. This drop indicates the energy release associated with the change from a ferromagnetic to a paramagnetic phase. The thermal transition is further supported by the magnetic susceptibility ($\chi$) and inverse susceptibility ($\chi^{-1}$) data depicted in Figure 6(b). As temperature increases, $\chi$ decreases, reflecting the breakdown of magnetic ordering due to thermal agitation, while the non-linear behavior of $\chi^{-1}$ emphasizes the complex nature of the magnetic phase transition occurring near $T_C$. Complementing these observations, Figure 6(c) reveals a sharp decrease in magnetization as the temperature approaches $T_C$, clearly marking the transition to paramagnetic state. This behaviour aligns with the earlier discussion on the short-range magnetic interactions and the noncollinear magnetic structure within the compound. Collectively, these figures offer a comprehensive understanding of how thermal energy influences magnetic ordering in β-IrMnAl, highlighting the crucial role temperature plays in determining its magnetic properties.

## 4. Conclusion

We have conducted a comprehensive investigation into the structural, elastic, electronic, and magnetic properties of IrMnAl, providing valuable insights into its phase transition behaviour, elastic moduli, and magnetic characteristics, particularly under pressure. Our findings reveal that the α-phase of IrMnAl, characterized by a face-centered cubic (f.c.c) structure, undergoes a transition to a β-phase with a slightly reduced lattice constant when subjected to a pressure of 5.6 GPa. This phase transition is energetically favourable, as supported by analyses of the enthalpy and optimization curves.

The electronic and magnetic properties, explored through density of states (DOS) and partial density of states (PDOS) analyses, highlight the intricate electron correlation effects, especially in the β-phase. Here, the GGA+U method more accurately captures the electronic structure. It is noteworthy that while the inclusion of U does not induce half-metallic ferromagnetism, it aids in providing a precise explanation of their magnetic properties. The weak magnetism observed in β-IrMnAl is attributed to significant hybridization between Mn and Ir states, with the magnetic properties being highly sensitive to Mn-Mn exchange interactions. These interactions, which are predominantly short-ranged, vary with distance and directly influence the magnetic ordering and the compound's Curie temperature. Our thermal analysis of the β-phase reveals a Curie temperature of 243.4 K, indicating a transition from ferromagnetic to paramagnetic behaviour at this temperature. Overall, this study offers a detailed understanding of the behaviour of IrMnAl under varying conditions, with significant implications for its potential applications in the field of magnetic materials science.

**Acknowledgements**

The research is partially funded by the Ministry of Science and Higher Education of the Russian Federation as part of World-class Research Center program: Advanced Digital Technologies (contract No. 075-15-2022-312 dated 20.04.2022).

**A. Laref** acknowledges the Research Center of Female Scientific and Medical Colleges, Deanship of Scientific Research, King Saud University, Saudi Arabia for financial support.

**Conflict of Interest**

The authors have no conflict to declare.